\documentclass[12pt]{article}
\usepackage{amssymb}
\usepackage{epsfig}

\newcommand{\bq}{\begin{equation}}
\newcommand{\eq}{\end{equation}}
\newcommand{\bqa}{\begin{eqnarray}}
\newcommand{\eqa}{\end{eqnarray}}
\newcommand{\ben}{\begin{enumerate}}
\newcommand{\een}{\end{enumerate}}
\newcommand{\bc}{\begin{center}}
\newcommand{\ec}{\end{center}}

\def\gsim{\gtrsim}
\def\lsim{\lesssim}

\def\pr#1#2#3{ Phys. Rev. ${\bf{#1}}$ (#2) #3}

\def\pl#1#2#3{ Phys. Lett. ${\bf{#1}}$ (#2) #3}

\def\np#1#2#3{ Nucl. Phys. ${\bf{#1}}$ (#2) #3}

\global\nulldelimiterspace = 0pt

\def\O{ {\cal O }}

\begin{document}
\thispagestyle{empty}
\begin {flushleft}

 PM/97-29\\

August 1997\\
\end{flushleft}

\vspace*{2cm}

\begin{center}
{\Large\bf  Identification of Virtual New Physics Effects   }\\
\vspace {0.1cm}
{\Large\bf  at a Linear Collider} \\

 \vspace{1.8cm}
{\large F.M. Renard$^a$ and C.
Verzegnassi$^b$}
\vspace {1cm}  \\

$^a$Physique
Math\'{e}matique et Th\'{e}orique,
UPRES-A 5032,\\
Universit\'{e} de Montpellier II,
 F-34095 Montpellier Cedex 5.\\
\vspace{0.2cm}
$^b$ Dipartimento di Fisica,
Universit\`{a} di Lecce \\
CP193 Via Arnesano, I-73100 Lecce, \\
and INFN, Sezione di Lecce, Italy.\\

\vspace{1.5cm}

 {\bf Abstract}
\end{center}
\noindent

It is shown that, at a 500 GeV LC, a number of different theoretical
models would be unambigously identified by their virtual effects.
Negative limits in case of no signal identification are also derived.

\vspace{3cm}

{\bf Contribution to the ECFA/DESY Study on 
Physics and Detectors for the Linear Collider.}\\
Partially supported by the EC contract CHRX-CT94-0579.

\vspace{1cm}

\setcounter{page}{0}
\def\thefootnote{\arabic{footnote}}
\setcounter{footnote}{0}
\clearpage

\section{Introduction}\par
\hspace{1cm}When the future 500 GeV linear 
electron positron collider, to be called
LC from now on, will become operative, several outstanding problems in
the field of elementary particles physics will have been, hopefully,
solved or at least understood. In particular, possible discoveries of
one or of several Higgs bosons and of SUSY particles at LEP2, Tevatron
and LHC might occur. Depending on these results, a number of orthogonal
theoretical models would necessary die. For example, Higgs production
would be the end of technicolour ideas, while models with anomalous
gauge couplings in conventional formulations would not survive the
discovery of supersymmetry. The latter might, in turn, be in serious
trouble if a "too" heavy Higgs were discovered, and this might cause
difficulties to a number of theoretical models involving one extra Z
imbedded in a SUSY scenario.\par

This preliminary short and qualitative discusssion has only shown that,
at the time of LC future runs, three classes of theoretical models,
that are still widely considered today as a possible alternative to a
conventional "minimal"
scheme that includes supersymmetry as well, might well have been
completely or partially discarded. We shall refer to these models as to
models of "technicolour type" (TC), models with anomalous gauge
couplings (AGC) and models with one 
extra $Z (\equiv Z')$ respectively.\par

Keeping the previous remarks in mind, we shall proceed from now on 
assuming that all these three classes of models will still be
meaningful at LC (for example, this might happen 
if no Higgs or SUSY particles were produced,
with surviving windows in the low Higgs mass region). The following
discussion will then be devoted to the possibility of identifying these
competitor models from their \underline{virtual effects} at LC in the
extremely plausible hypothesis that no kind of direct production is
alternatively allowed. It would be relatively simple to mofify the
discussion if some of the models became inadequate when LC will begin
to run, or, also, if new types of models were at that time proposed.\par

The simple idea that is at the basis of the following discussion is the
fact that, for any theoretical model with a suficiently low number of
parameters $\equiv N$, a relationship exists between the relative
effects of the model on certain suitable $N+1$ experimental
observables. This is fixed by the model, but at the same time
completely free of the (arbitrary) N parameters. The relationship can
be in fact easily obtained by eliminating the parameters i.e.
expressing them in terms of the considered relative effects. As a
result of this operation, one can draw a certain figure in the
(N+1)-dimensional space of the effects that is characteristic of the
considered model.\par

An important example of the previous vague statement is provided by
models with two free parameters. The corresponding three-dimensional
figures that the models generate were called in a previous paper
\cite{R1} "reservations". Clearly, to produce such figures,
\underline{three} experimental observables are requested.\par

Let us consider now the specific cases that were
mentioned at the beginning. It was shown 
in previous references \cite{R2}
that for a simple class of models of "technicolour-type" a
representation of four-fermion (neutral current) processes can be
written, such that the effects of these models can be represented by
two effective parameters. This special representation was called
"$Z$-peak subtracted representation"; we defer the interested reader to
ref.\cite{R2} for more details. Therefore, for the previous class of TC
models. three (arbitrary) observables of the process $e^+e^-\to f\bar
f$ ($ f$ being a charged lepton or a quark) will be sufficient to draw
the TC reservation.\par

The case of AGC models was also considered in ref \cite{R2}, taking as
specific example that of a $SU(2)\times U(1)$ invariant Lagrangian
with only dimension six operators recently proposed \cite{R3}. Although
the number of parameters of this model is apparently rather large, it
was shown in ref.\cite{R2} that, in the $Z$-peak subtracted
representation, only \underline{two} parameters survive in the
considered four-fermion process (with the possible exception of final
$b\bar b$ production). This would lead to possible 3d-figures,
reservations for this model, as well.\par

The case of a possible $Z'$ requires some care. In general, if the
couplings with fermions of this extra $Z$ are left free, for a process
like $e^+e^-\to f\bar f$ \underline{six} effective parameters will
enter, as discussed e.g. in a recent investigation \cite{R4}. This
number can be reduced to \underline{two} if only a final (charged)
lepton-antilepton state is considered.\par

From the previous discussion it emerges that, if \underline{three}
suitable independent observables were available in the process
$e^+e^-\to f\bar f$, one would be able to draw reservations for the
three different models here reviewed. If the respective reservations
did not overlap, it would be possible to conclude that the models would
be, in principle, clearly identifiable.\par

In practice, at LC with no polarized initial beams, one will be able to
measure with satisfactory accuracy \cite{R5} in the final lepton
channel the lepton 
(typically, muon, tau) cross section $\sigma_l$ and the
related forward-backward asymmetry $A_{FB,l}$. These two observables
would not be sufficient to draw reservations. This would be possible
if a third independent leptonic observable were measured. The only
realistic possibility would be represented by the measurement of
$A_{LR,l}$, the longitudinal polarization asymmetry for production of
final leptonic states (at LC, this quantity would be quite different
from the corresponding hadronic one, not like at LEP/SLC). The
combination of $A_{LR,l}$ with $\sigma_l$, $A_{FB,l}$ would be
sufficient to draw typical regions of effects for each of the three
considered models.\par

If no initial state polarization were available, the most immediate
realistic choice would be provided by the measurement of the total
hadronic cross section, in particular by that for production of the
five "light" $(u,d,s,c,b)$ quarks ($\equiv \sigma_5$). The combined set
of $\sigma_{\mu}$, $A_{FB,l}$ and $\sigma_5$ would be sufficient to
draw reservations for the two TC and AGC models. In Section 2
we shall be limited to this
preliminary case.\par

\section{The unpolarized case.}\par

\hspace{1cm}This discussion can be made  
clearer by writing at this point a few
explicit formulae. In the $Z$-peak subtracted representation the
theoretical expressions of the considered observables read, at c.m.
squared energy $\equiv q^2$ :

1) the charged lepton pair production cross section
\bqa  \sigma_l(q^2)&=&\sigma^{Born}_{l}(q^2)\bigm\{1+{2\over
\kappa^2(q^2-M^2_Z)^2+q^4}[\kappa^2(q^2-M^2_Z)^2
\tilde{\Delta}_{\alpha}(q^2)\nonumber\\
&&-q^4(R(q^2)+{1\over2}V(q^2))]\bigm\} \eqa
\noindent
where $\kappa\equiv{\alpha M_Z\over3\Gamma_l}\simeq2.64$,
$\Gamma_l$ is the $Z$ width into $l^+l^-$ and
\bq
\sigma^{Born}_{l}(q^2)= {4\pi\alpha^2\over3q^2}
[{ q^4+\kappa^2(q^2-M^2_Z)^2\over\kappa^2(q^2-M^2_Z)^2}]      \eq

\noindent

2) its associated forward-backward asymmetry
\bqa  A_{FB,l}(q^2)&=&A^{Born}_{FB,l}(q^2)\bigm\{1+
{q^4-\kappa^2(q^2-M^2_Z)^2
\over\kappa^2(q^2-M^2_Z)^2+q^4}[
\tilde{\Delta}_{\alpha}(q^2)+R(q^2)]\nonumber\\
&&+{q^4\over\kappa^2(q^2-M^2_Z)^2+q^4}V(q^2)]\bigm\} \eqa
\noindent
where
\bq
A^{Born}_{FB,l}(q^2)= {3q^2\kappa(q^2-M^2_Z)
\over2[ q^4+\kappa^2(q^2-M^2_Z)^2]}      \eq
\noindent

3) the hadronic cross section 
$\sigma_5(q^2)$ ($\equiv
\sigma_u(q^2)+\sigma_d(q^2)+\sigma_s(q^2)+\sigma_c(q^2)
+\sigma_b(q^2)$)
for which a relatively simple approximate expression
has been written \cite{R7}

\bqa  \sigma_5(q^2)\simeq &&N^{QCD}_q(q^2){4\over3}
\pi q^2\{[{11\alpha^2\over
9q^4}][{1\over1-\tilde{\Delta}_{\alpha}(q^2)}]^2+\nonumber\\
&&[({3\Gamma_l\over M_Z})({3\Gamma_{had}\over
M_Z N^{QCD}_q(M^2_Z)})({1\over
q^2-M^2_Z})][1-2R(q^2)-s_l c_l V(q^2)({32\over10}{
\Gamma_c\over\Gamma_{had}}+{48\over13}{\Gamma_b\over\Gamma_{had}})]\}
\nonumber\\
&&+[({2\alpha\over q^2})({2\tilde{v}_l\over
q^2-M^2_Z})\sqrt{({3\Gamma_l\over M_Z})({3\Gamma_{had}\over
M_ZN^{QCD}_q(M^2_Z)})}({2\over9}\sqrt{{
\Gamma_c\over\Gamma_{had}}}+{3\over\sqrt3}
\sqrt{{\Gamma_b\over\Gamma_{had}}})\}
\label{2-31}\eqa
where $\Gamma_c$, $\Gamma_b$, $\Gamma_{had}$ are the $Z$ widths into 
$c\bar c$, $b\bar b$ and hadrons, and $\tilde{v}_l\equiv1-4s^2_l$
where $s^2_l=1-c^2_l$ is the
effective Weinberg-Salam parameter measured on $Z$ resonance.

The "Born" terms consist in fact of expressions where the Fermi
constant, $G_{\mu}$, has been systematically traded for quantities
measured on $Z$ resonance (in the case of lepton production,
 eqs.(1)-(4), the
leptonic Z width $\Gamma_l$ and the longitudinal polarization asymmetry
at $Z$ peak $A(M^2_Z)$ are involved). 
One sees that the shifts from the Standard
Model prediction will be contained in three functions
$\tilde{\Delta}_{\alpha}(q^2)$, $R(q^2)$, $V(q^2)$.\par

Their expressions in the three considered models contain in each case
two parameters. The explicit formulae have been given in
ref.\cite{R1} and \cite{R2}. In
first approximation, they are rather simple. We consider first the case
of the models with AGC. Here one finds, (neglecting irrelevant small
contributions):

\bq \tilde{\Delta}^{(AGC)}_{\alpha}(q^2)=
-8\pi\alpha{q^2\over\Lambda^2}
(f^r_{DW}+f^r_{DB}) \eq

\bq  R^{(AGC)}(q^2)= 8\pi\alpha{(q^2-M^2_Z)\over\Lambda^2}
({c^2_1\over s^2_1}f^r_{DW}+{s^2_1\over c^2_1}f^r_{DB}) \eq

\bq V^{(AGC)}(q^2) = 8\pi\alpha{(q^2-M^2_Z)\over \Lambda^2}
({c_1\over s_1}f^r_{DW}-{s_1\over c_1}f^r_{DB})  \eq

\noindent
where $f^r_{DW}$, $f^R_{DB}$ are the renormalized parameters that, in
the considered example, are associated to the operators
$\O_{DW}$ and $\O_{DB}$ involved in the effective Lagrangian of
ref.\cite{R3} 
and that survive in the Z-peak subtracted
representation. As one sees, the previous  equations are linear in the
two surviving parameters of the model. Eliminating the parameters leads
to a very simple figure in the 3d-space of the relative deviations that
is , in fact, that of a plane.\par

The result of this procedure is shown in Fig.(1), that represents, in
the space of the three relative shifts on $\sigma_l$, $A_{FB,l}$ and
$\sigma_5$, the corresponding reservations at $\sqrt{q^2}=500~GeV$. The
central box represents the region of non visibility of a signal,
fixed by the expected experimental accuracy at LC discussed in
ref.\cite{R5}.\par

Fig.(1) (light grey domain) shows the region 
of a certain space where a signal of AGC
origin would be identifiable, at least as a possible candidate. In the
negative case of no signal detection, limits on the two previous
parameters would be derivable from the combined analysis of the three
considered observables. This problem has been fully discussed in a
recent paper \cite{R6}, where the potentially dangerous effect of
(initial state) QED radiation have been also taken into account. The
resulting exclusion regions for the parameters
are shown in Fig.(2). Numerically, they can be summarized, at 95\% CL,
by the following limits:

\bq
|f^r_{DW}|\lsim0.025
\eq

\bq
|f^r_{DB}|\lsim0.13
\eq

We now move to an example of a model of TC type. We shall  choose a
case in which \underline{two} families of strong vector and
axial-vector techniresonances exist, that would contribute the neutral
gauge bosons self-energies and thus the coefficients
$\tilde\Delta_{\alpha}$, $R$ and $V$. This has been exhaustively discussed
in a quite recent paper \cite{R7}, to which we defer for details. In
the particularly simple (but realistic) case in which for each family
one "light" resonance exists, such that the next resonances are
"reasonably" heavier, and assuming a certain self-consistent size of
the strengths of their couplings, this model can be also described by
two $q^2$-dependent parameters, defined as:

\bq  X(q^2)= ({F^2_V\over M^2_V})({1\over M^2_V-q^2})
\label{2-39}\eq

\bq  Y(q^2)= ({F^2_A\over M^2_A})({1\over M^2_A-q^2})
\label{2-40}\eq

In terms of $X$, $Y$ the \underline{relative} shifts of the three
considered observables are :

\bq  {\delta\sigma^{(TC)}_{\mu}\over \sigma_{\mu}}(q^2)=
{\sigma^{(TC)}_{\mu}-\sigma^{(SM)}_{\mu}\over \sigma_{\mu}}=
a_1(q^2) X(q^2)+b_1(q^2) Y(q^2)
\label{2-41}\eq

\bq   {\delta A^{(TC)}_{FB,\mu}\over A_{FB,\mu}}(q^2)=
{A^{(TC)}_{FB,\mu}-A^{(SM)}_{FB,\mu}\over A_{FB,\mu}}=
a_2(q^2) X(q^2)+b_2(q^2) Y(q^2)
\label{2-42}\eq

\bq   {\delta\sigma^{(TC)}_5\over \sigma_5}(q^2)=
{\sigma^{(TC)}_5-\sigma^{(SM)}_5\over \sigma_5}=
a_3(q^2) X(q^2)+b_3(q^2) Y(q^2)
\label{2-43}\eq

\noindent
where $a_i$, $b_i$ are \underline{numbers} whose value at LC is
calculable, and $F_V$, $F_A$ are the resonances' strengths,
conventionally defined.\par

Eqs.(13)-(15) generate the corresponding TC reservation. At 500 GeV,
this is depicted in Fig.(1) (dark grey domain), 
where the same conventions used in the case of AGC
have been adopted.\par

As one sees from a comparison of the domains in Figs.(1), 
the overlapping
region between the two considered TC and AGC models is practically
negligible. As a consequence, a virtual signal due to
a model of the first type would be clearly distinguishable, at LC, from
a model of the second type by a combined use of the three unpolarized
onservables that will be realistically measured with the expected
accuracy.\par

As it was done for the first case of AGC models, negative bounds for
the TC parameters in case of no observation of the related signal can
be drawn. The discussion has been fully pursued in Ref.\cite{R7},
leading to the following mass \underline{exclusion} lower limits for
the vector ($M_V$) and axial vector ($M_A$) resonance in case of a
typical reasonable choice of the ($F_V$, $F_A$) strengths:

\bq
M_V\gsim1.5 ~TeV
\eq

\bq
M_A\gsim0.9~TeV
\eq

In this discussion, we have assumed the non availability of initial
longitudinal polarization. We shall examine in the next Section 3
the possible consequences of the opposite situation of
polarization availability.\par

\section{The polarized case.}\par
\hspace{1cm}
Among the several quantities that can be measured in the process of
electron-positron annihilation into a fermion-antifermion couple, the
longitudinal polarization asymmetry $A_{LR} \equiv
{\sigma_{L}-\sigma_{R}\over \sigma_{L}+\sigma_{R}}$ has
represented in the last few years an example of, least to say,
remarkable theoretical interest. This is due to the known fact that, as
it was stressed in a number of dedicated papers \cite{R8},\cite{R9},
\cite{R10},\cite{R11}, the properties of this observable on top of $Z$
resonance are indeed special. In particular one can
stress two main facts i.e. that $A_{LR}$ is independent of the final
produced state (this was shown in particular detail in Ref.\cite{R10}),
and that it is particularly sensitive to possible virtual effects of a
large number of models of new physics (this was exhaustively discussed
in Refs.\cite{R9} and \cite{R10}). 
These features, that appear essentially
unique, have deeply motivated the tough experimental effort at SLC
\cite{R12} where $A_{LR}$ has been (in fact, it is still being)
measured to an extremely high precision \cite{R13}, fully exploiting the
fact that at a linear electron-positron collider it is "relatively"
easy to produce longitudinally polarized electron beams with a high and
accurately known polarization degree\cite{R14}. 
This is not the case of a
circular accelerator, and for this reason neither at LEP1 (in spite of
the several impressive experimental studies and efforts of recent years
\cite{R15}) nor at LEP2 a measurement of $A_{LR}$ has been, or will be
predicticably performed.\par

 At LC it would be, again,
"relatively" easy to produce longitudinally polarized electron beams,
which implies the possibility of measuring $A_{LR}$, for
various possible final states. One might therefore wonder whether the
special theoretical properties valid on top of Z resonance will still
be true and, if not, how they would be modified at about 500 GeV.
This question has been answered in a very recent paper \cite{R16} which
has investigated the general features of $A_{LR}$ at such a machine,
showing that, from a theoretical point of view, this quantity still
retains beautiful and interesting features, that make it particularly
promising as a tool for investigating virtual effects of models of new
physics. We defer the interested reader to ref.\cite{R16} for more
details, and proceed by giving the theoretical expression of $A_{LR}$
that will be relevant for our purposes. In particular, from the
discussion given in Section 1, we shall concentrate our attention on
the final lepton states, for which the Z-peak subtracted representation
of $A_{LR,l}$ reads, at \underline{one loop} (specified by the notation
$A^{(1)}_{LR,l}$):

\bqa &&A^{(1)}_{LR,l}(q^2)={q^2[\kappa(q^2-M^2_Z)+q^2]
\over \kappa^2(q^2-M^2_Z)^2+q^4}A_{LR}(M^2_Z)\times
\nonumber \\
&&\bigm\{1+[{\kappa(q^2-M^2_Z)
\over\kappa(q^2-M^2_Z)+q^2}-{2\kappa^2(q^2-M^2_Z)^2\over
\kappa^2(q^2-M^2_Z)^2+q^4}]
[\tilde{\Delta}_{\alpha}(q^2)+R(q^2)]
-{4c_ls_l\over \tilde{v}_l}V(q^2) \bigm\} \eqa

\noindent
where $\kappa$ has been defined after eq.(1).

To make practical use of this extra information, we shall assume an
experimental accuracy in the measurement of $A_{LR,l}$ at LC of about
$0.007$ (purely statistical), to be compared with the expected value
$A_{LR,l}(500~GeV)\simeq0.07$. All our results can be easily rescaled
if the experimental error is raised, or decreased. As a first example,
we shall consider the case of a model with one 
extra $Z (\equiv Z')$. For a
final lepton state, this case can be classified in the $Z$-peak
subtracted framework as one described by two parameters. More
precisely, as it has been shown in a previous dedicated review
\cite{R4}, one finds

\bq \tilde{\Delta}^{(Z')}_{\alpha}(q^2)=-{q^2\over M^2_{Z'}-q^2}({1\over
4c^2_l s^2_l})g^2_{Vl}(\xi_{Vl}-\xi_{Al})^2
\eq

\bq    R^{(Z')}(q^2)=({q^2-M^2_Z\over M^2_{Z'}-q^2})\xi_{Al}^2
\eq

\bq  V^{(Z')}(q^2)=-({q^2-M^2_Z\over
M^2_{Z'}-q^2}){g_{Vl}\over
2c_l s_l}\xi_{Al}(\xi_{Vl}-\xi_{Al})
\eq
\noindent
where we have used the definitions

\bq  \xi_{Vl}={g'_{Vl}\over g_{Vl}} 
\eq

\bq \xi_{Al}={g'_{Al}\over g_{Al}} 
\eq
 \noindent
with $g_{Vl}={1\over2}(1-4s^2_l)$; $g_{Al}=-{1\over2}$ and
$g'_{Vl}$, $g'_{Al}$, the (arbitrary) Z' couplings to
leptons.\par
As one sees, only two effective parameters, that could be taken for
instance as\\ 
$\xi_{Vl}{M_Z\over\sqrt{M^2_{Z'}-q^2}}$ and 
$(\xi_{Vl}-\xi_{Al}){M_Z\over\sqrt{M^2_{Z'}-q^2}}$,
 enter the one-loop corrections $\tilde{\Delta}^{(Z')}_{\alpha}(q^2)$, 
$R^{(Z')}(q^2)$ and $V^{(Z')}(q^2)$. This
means that, as stated in Section 1, a proper combination of the
theoretical expressions of the three leptonic quantities $\sigma_l$,
$A_{FB,l}$ and $A_{LR,l}$ allows to
derive the characteristic Z' reservation. This is shown in Fig.(3) for
the LC case.\par
Assuming that $A_{LR,l}$ can be measured, we can reconsider the two
previous models of TC and AGC type examined in Section 1 and draw their
reservations in the space of the three leptonic shifts observables 
$\delta\sigma_l$,
$\delta A_{FB,l}$ and $\delta A_{LR,l}$
This is shown in Fig.(4). Then, a comparison of the different
characteristic areas for the \underline{three} models of TC, AGC and Z'
type can be performed. As one sees from Fig.(3),(4), the common region
to the three models is, again, very small. A
nice consequence of the availability of initial beam polarization would
be thus the possibility of disentangling in a clean way the possible
effects of \underline{three} respectable models of New Physics of
rather different nature.\par

One should still add the fact that, for what concerns the Z' model,
limits on its mass from absence of signals 
have been derived both in the presence and
in the absence of polarization. They are summarized in other previous
papers\cite{R17}, and lie typically in the few TeV range at LC.
 We give, for illustration purposes, in Table 1,
 a few numerical values (with only statistical errors), 
corresponding to special
"canonical" models, assuming the presence of polarization (this
improves the bounds  by factors that can be sizeable, depending on the
model).\par

\section{Conclusions.}

\hspace{1cm}In this short paper we have reviewed  
the main information that would be
obtainable at LC, from a proper combined use of different observables,
concerning three rather general and quite different models of New
Physics. Our analysis shows that it would be possible to identify the
possible origin of a signal with reasonable accuracy, particularly if
longitudinal polarization were available for the 
initial electron beam. We have
also derived limits on the "effective" parameters in the Z peak
subtracted approach, assuming the absence of visible signals, that in
the case of vector resonance masses lie typically in the TeV range. Our
analysis can be easily generalized to the case of a similar machine with
different c.m. energy, without changing any of the relevant features of
our approach.

\vspace{0.5cm}
\leftline{\Large \bf Acknowledgments}
We thank Jacques Layssac for his help in the preparation of the
3-d figures.
\newpage

\vspace{1cm}
\begin{center}
{\bf Table 1: $Z'$ mass limits in left-right and $E_6$ models.}

\vspace{1cm}
\begin{tabular}{|c|c|c|c|c|} \hline
\multicolumn{1}{|c|}{$\alpha_{LR}$}&
\multicolumn{1}{|c|}{$0.8$} &
\multicolumn{1}{|c|}{$1.$}&
\multicolumn{1}{|c|}{$1.2$}&
\multicolumn{1}{|c|}{$1.4$} 
 \\[0.1cm] \hline
&&&&\\
$M_{Z'}$ (TeV)&$4.2$&$3.1$&$3.1$&$3.8$\\
 \hline
\end{tabular}
\begin{tabular}{|c|c|c|c|c|c|} \hline
\multicolumn{1}{|c|}{$cos\beta_{E_6}$}&
\multicolumn{1}{|c|}{$-1$} &
\multicolumn{1}{|c|}{$-0.5$}&
\multicolumn{1}{|c|}{$0.$}&
\multicolumn{1}{|c|}{$0.5$}&
\multicolumn{1}{|c|}{$1$}\\[0.1cm] \hline  
&&&&&\\
$M_{Z'}$ (TeV)&$4.$&$2.1$&$1.8$&$3.2$&$4.1$
\\ \hline
\end{tabular}
\end{center}

\newpage
\begin{center}

{\large \bf Figure captions}\\
\end{center}
\vspace{0.5cm}

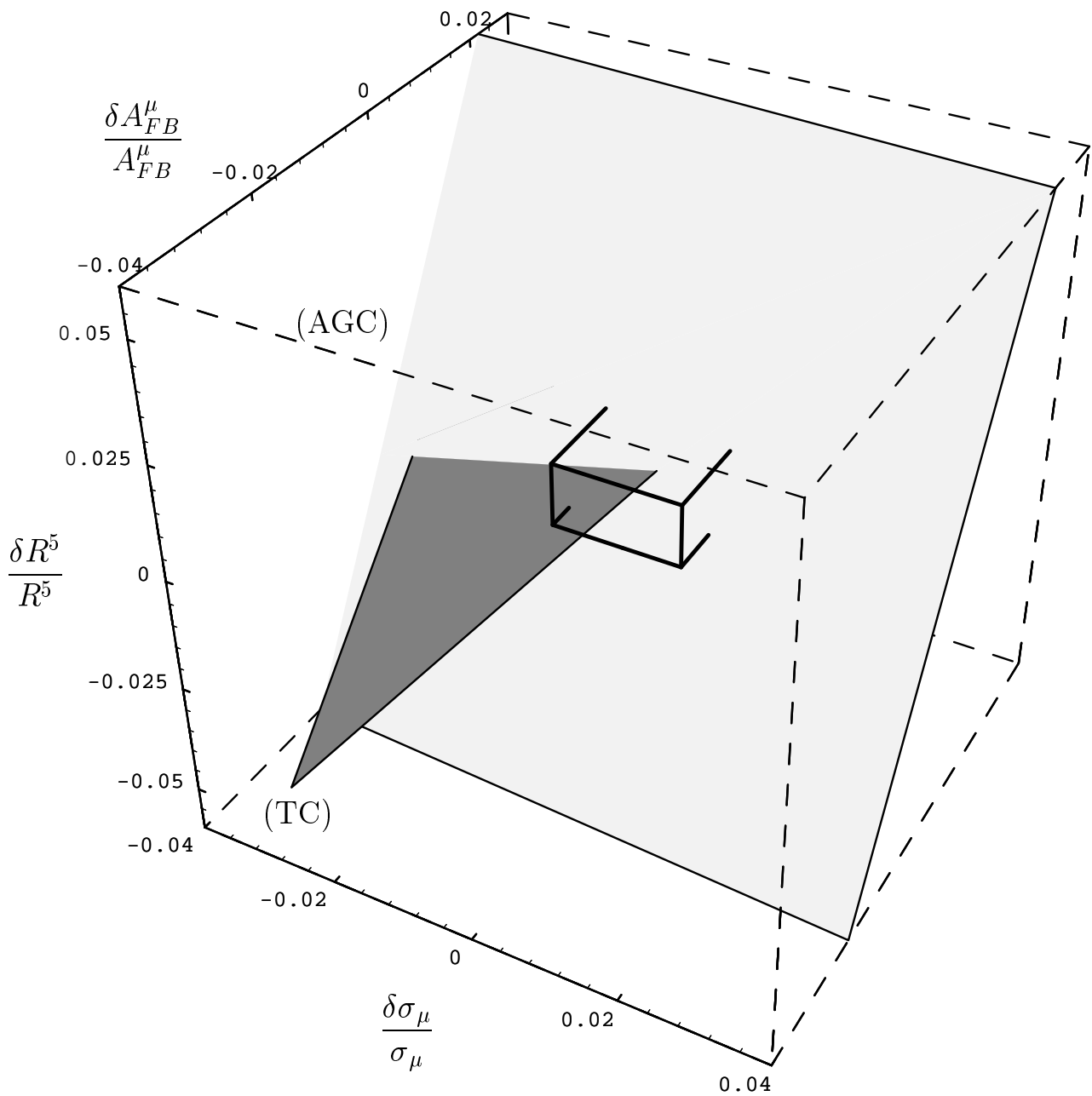
{\bf Fig.1}. Reservations  for AGC models (light grey) and TC
models (dark grey) in the space of relative shifts on 
$\sigma_l$, $A_{FB,l}$ and $\sigma_5$ at a 500 GeV LC. The central
box represents the unobservable domain.\\

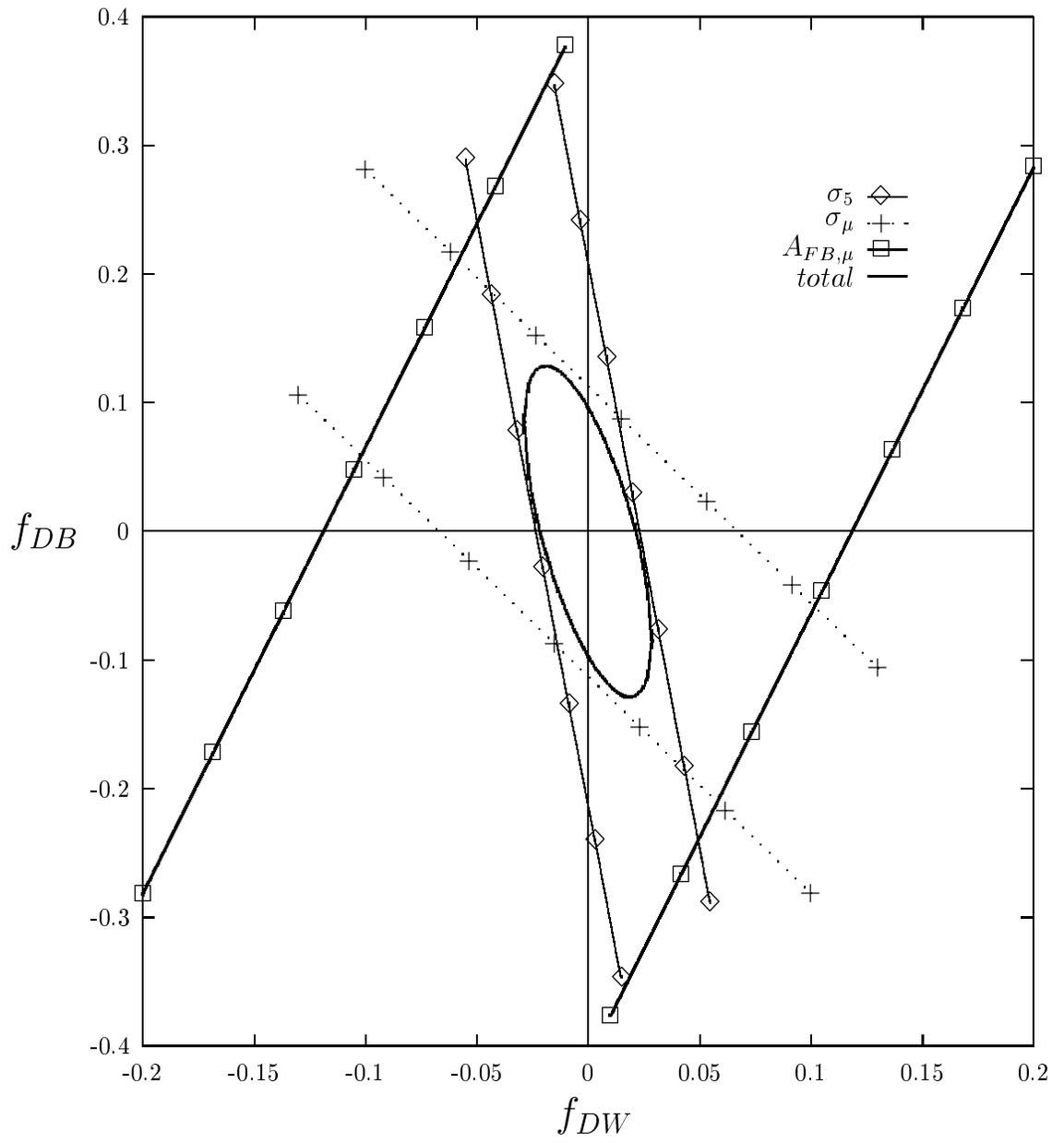
{\bf Fig.2} Constraints on AGC couplings from $e^+e^-\to f\bar f$ 
processes at 500 GeV without polarization.
$\sigma_l$ ($cross$), $\sigma_5$ ($diamond$),
$A_{FB,l}$ ($box$).
The ellipse is
obtained by combining quadratically the three bands. \\ 

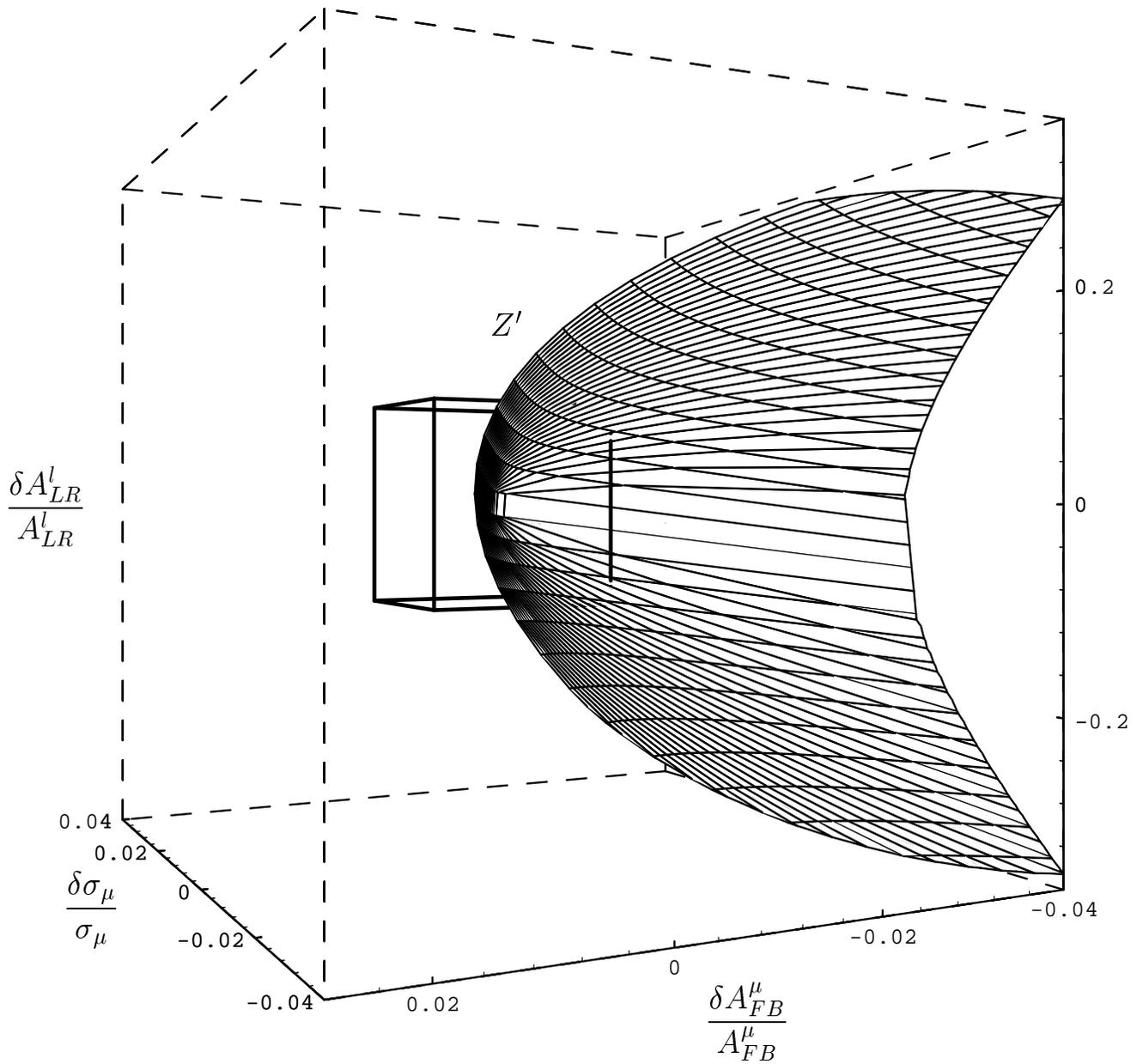
{\bf Fig.3}.  Reservations  for general $Z'$ models in the 
space of relative shifts on 
$\sigma_l$, $A_{FB,l}$ and $A_{LR,l}$ at a 500 GeV LC.\\

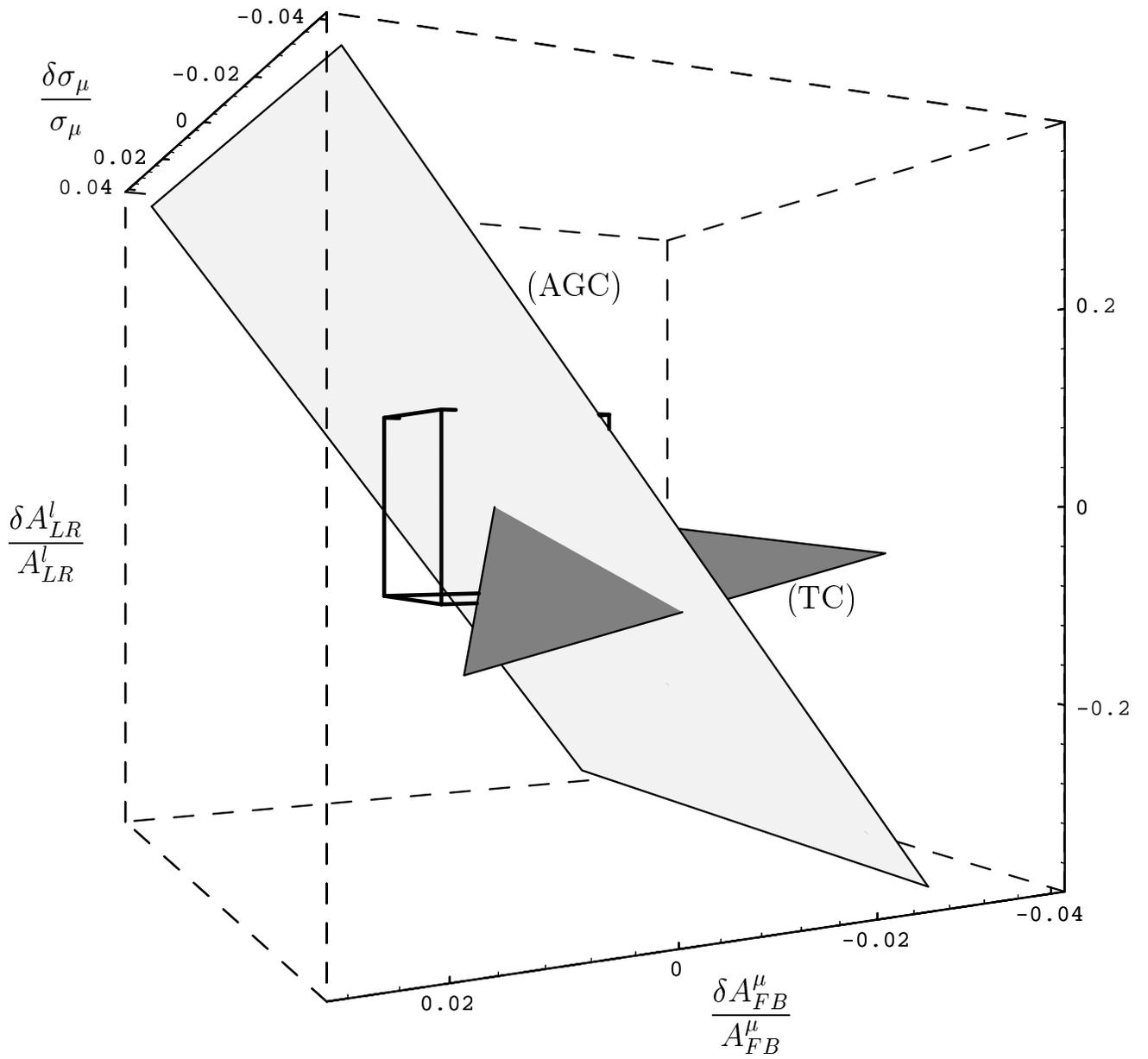
{\bf Fig.4}.  Reservations  for AGC models (light grey) and TC
models (dark grey) in the 
space of relative shifts on 
$\sigma_l$, $A_{FB,l}$ and $A_{LR,l}$ at a 500 GeV LC.\\

\end{document}